\documentclass[9pt,twocolumn,letterpaper]{article}
\usepackage[utf8]{inputenc}
\usepackage[english]{babel}
\usepackage{amsmath,amsfonts,amssymb}
\usepackage{lmodern}
\usepackage[scaled]{helvet}
\usepackage[T1]{fontenc}
\usepackage[normalem]{ulem}
\usepackage{authblk}
\usepackage[numbers,sort&compress,merge,round]{natbib}
\usepackage{graphicx,xcolor}
\usepackage[margin=0.5in]{geometry}
\title{Epsilon-iron as a spin-smectic state} 

\author[a,b,1]{Blair W. Lebert}
\author[a,1]{Tommaso Gorni}
\author[a]{Michele Casula}
\author[a]{Stefan Klotz}
\author[b]{Fran\c{c}ois Baudelet}
\author[b]{James M. Ablett}
\author[c]{Thomas C. Hansen}
\author[a]{Am\'{e}lie Juhin}
\author[a,b]{Alain Polian}
\author[a]{Pascal Munsch}
\author[d]{Gilles Le Marchand}
\author[a]{Zailan Zhang}
\author[b,e]{Jean-Pascal Rueff}
\author[a,f,2]{Matteo d'Astuto}

\affil[a]{IMPMC, UMR CNRS 7590, Sorbonne Universit\'es-UPMC University, MNHN, 4 Place Jussieu, F-75005 Paris, France}
\affil[b]{Synchrotron SOLEIL, L'Orme des Merisiers, BP-48 Saint-Aubin, 91192 Gif-sur-Yvette, France}
\affil[c]{Institut Laue-Langevin, Grenoble, France}
\affil[d] {IGDR, UMR 6290 CNRS -UR1, Campus Sant\'e de Villejean, Facult\'e de M\'edecine, 2 Avenue du Professeur L\'eon Bernard, 35043 Rennes Cedex}
\affil[e]{Sorbonne Universit\'e, CNRS, Laboratoire de Chimie Physique - Mati\`ere et Rayonnement, LCPMR, 75005 Paris, France}
\affil[f] {Univ. Grenoble Alpes, CNRS, Grenoble INP, Institut N\'eel, 38000 Grenoble, France}

\begin{document}
\maketitle
\footnotetext[1]{B.W.L and T.G contributed equally to this work}
\footnotetext[2]{Corresponding author - matteo.dastuto@neel.cnrs.fr - Institut N\'{e}el CNRS - 25, av des Martyrs - 38042 Grenoble cedex 9; tel: (+33)(0)4 76 88 12 84}
\begin{abstract}
Using x-ray emission spectroscopy, we find appreciable local magnetic moments until 30--40\,GPa in the high-pressure phase of iron, however no magnetic order is detected with neutron powder diffraction down to 1.8\,K contrary to previous predictions. Our first-principles calculations reveal a ``spin-smectic'' state lower in energy than previous results. This state forms antiferromagnetic bilayers separated by null spin bilayers, which allows a complete relaxation of the inherent frustration of antiferromagnetism on a hexagonal close-packed lattice. 
The magnetic bilayers are likely orientationally disordered, owing to the soft interlayer excitations and the near-degeneracy with other smectic phases.
This possible lack of long-range correlation agrees with the null results from neutron powder diffraction. An orientationally-disordered, spin-smectic state resolves previously perceived contradictions in high pressure iron and could be integral to explaining its puzzling superconductivity.
\end{abstract}

\begin{figure}[!htb]
\centering
\includegraphics[width=\linewidth]{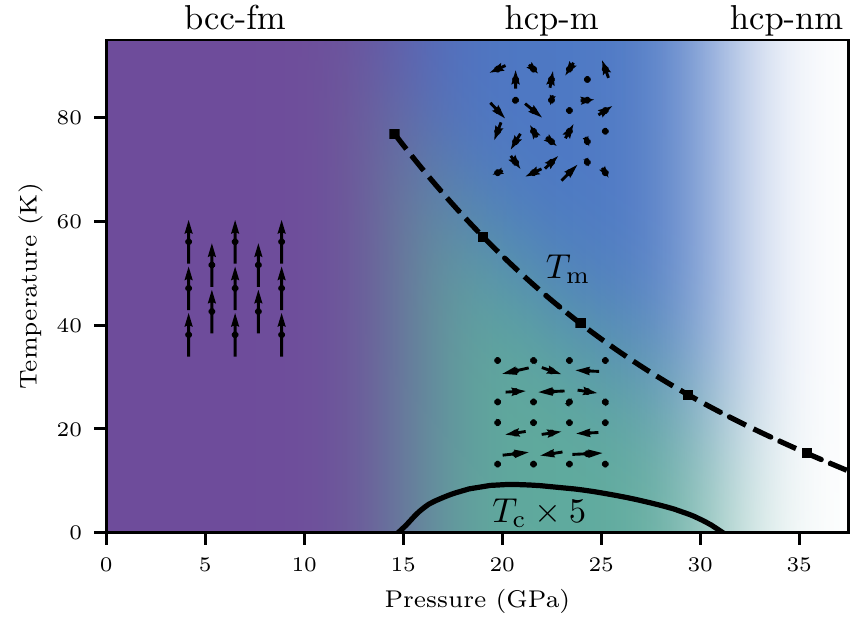}
\caption{Schematic phase diagram of iron. The ambient ferromagnetic body-centered cubic phase (bcc-fm) is shown in purple. The high-pressure form $\varepsilon$-iron, which forms a hexagonal close-packed (hcp) structure, is shown in blue/green for magnetic phases (hcp-m) and white for non-magnetic phases (hcp-nm). Our calculations predict that the disordered moments (blue) form a spin-smectic state (green) below a critical temperature $T_{\rm m}$, which may be related to $\varepsilon$-iron's superconductivity ($T_{\rm c}$ dome shown).}
\label{fig:phase}
\end{figure}

Iron is well-known since antiquity for its unique magnetic properties and continues to captivate scientists to this day. The study of iron and its alloys has many applications, including steel production and geophysics. Regarding the latter, the application of hydrostatic pressure induces a phase transition from the body-centered cubic (bcc) structure of $\alpha$-iron to the hexagonal close-packed (hcp) structure of $\varepsilon$-iron (Fig.~\ref{fig:phase}). Iron is being studied at increasingly high pressures and temperatures, since it and its alloys compose the majority of the Earth's core 
\cite{Jeanloz1990}.
Nonetheless, the relatively low-pressure, low-temperature region of $\varepsilon$-iron has remained a mystery for decades. The ferromagnetism (fm) found in $\alpha$-iron disappears during the $\alpha$-$\varepsilon$ transition \cite{Mathon2004a, Baudelet2005, Monza2011}, however the magnetic state of $\varepsilon$-iron is controversial. This became increasingly relevant after reports of unconventional superconductivity in this pressure range 
\cite{Shimizu2001}. 

K$\beta$ x-ray emission spectroscopy (XES) recently found a local magnetic moment \cite{Monza2011} in $\varepsilon$-iron, however magnetic order is undetected down to 30\,mK using M\"{o}ssbauer spectroscopy \cite{Cort1982}. Conversely, Raman spectroscopy observes mode splitting until 40\,GPa \cite{Merkel2000}, possibly from symmetry breaking by magnetic order. Density functional theory (DFT) calculations have predicted a collinear antiferromagnetic (afm) ground state, afmII, composed of alternating magnetization along the hcp $a$-axis 
\cite{Steinle-Neumann2004}
(Fig.~\ref{fig:DFT_phases}, right panel) which was consistent with null M\"{o}ssbauer spectroscopy results and Raman mode splitting. However, recent DFT calculations also predict an afmII state in the Fe\textsubscript{92}Ni\textsubscript{8} alloy, but with a substantial hyperfine magnetic field unlike in pure iron, yet synchrotron M\"{o}ssbauer spectroscopy still detects no magnetism \cite{Papandrew2006}. Furthermore, low-temperature Raman spectroscopy discovered that the splitting disappears contrary to expectations for magnetic order \cite{Goncharov2003}.

In this work, we performed K$\beta$ x-ray emission spectroscopy with considerably higher statistics, a different analysis technique, and a better pressure transmitting medium than past results \cite{Rueff1999a, Rueff2008a, Monza2011}. We confirmed that $\varepsilon$-iron has an intrinsic local moment \cite{Monza2011} and discovered that it decreases towards zero at 30--40\,GPa which is the same pressure range where its superconductivity disappears. We searched for possible ordering of these moments using neutron powder diffraction at record high-pressure and low-temperature conditions \cite{Klotz-HPR-NPD-Fe} and found no magnetic order down to 1.8\,K. The upper limit on the magnetic moment for afmII is five times smaller than theoretical estimates for the afmII configuration.

We searched for new magnetic phases in $\varepsilon$-iron using DFT and found spin-intensity-modulated (im) phases (Fig.~\ref{fig:DFT_phases}, right panel) lower in energy than the afmII phase in the pressure range where local magnetic moments are experimentally detected. We show that spin-intensity modulation is favored by the lattice frustration and the large spin degeneracy of iron sites in the hcp geometry. In order to account for this modulation, we derive an extended Heisenberg model from first principles, where spins are allowed to vary in both direction (transverse fluctuations) and magnitude (longitudinal fluctuations). Below a critical temperature (of about 
55\,K
at 20GPa), the finite-temperature solution of this model ---~sampled by classical Monte Carlo~--- consists of spatially separated afm bilayers, found also by our DFT calculations at zero temperature (imIII phase). This is a ``spin-smectic'' arrangement since the longitudinal fluctuations break the lattice translation symmetry much in the same way a liquid crystal in the smectic phase breaks the translation invariance of a given direction 
\cite{stephen1974}. 
The spin-smectic arrangement perfectly cancels the spin frustration in the hcp antiferromagnetic lattice. While every bilayer is antiferromagnetically ordered, the orientational order between bilayers crucially depends on the spin coupling beyond nearest neighbors. 
We found the next-nearest-neighbors (interlayer) coupling 
to drop from $2$ meV/$\mu_B^2$ to $0.2$ meV/$\mu_B^2$
in the 20-35\,GPa pressure range.
The combined effect of the spin-flip softness due to a progressively weaker 
interlayer coupling, together with the proximity with other smectic phases, could lead to fluctuations destroying long-range order.
This picture 
is consistent with our x-ray emission spectroscopy and neutron powder diffraction results, as well as previous experimental findings.

\begin{figure}[!t]
\centering
\includegraphics[width=\linewidth]{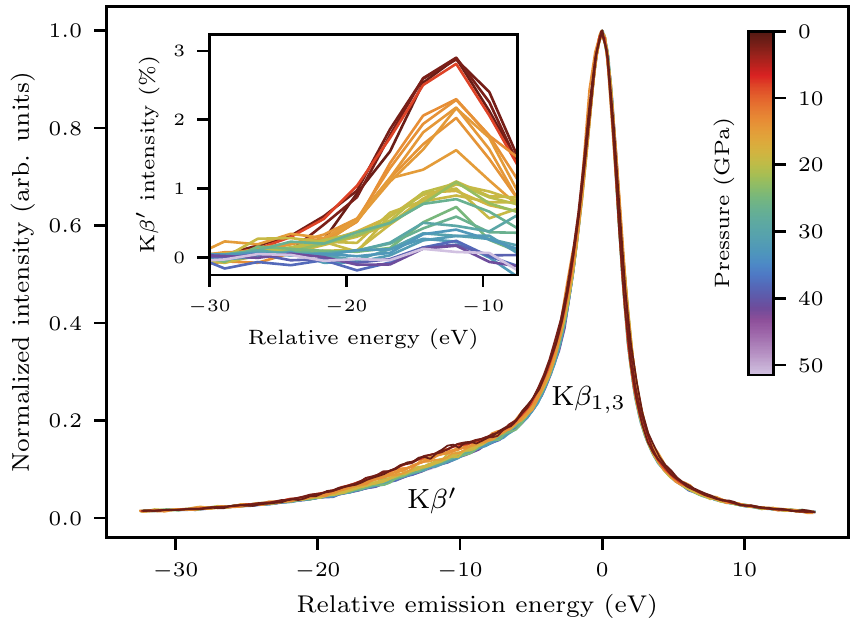}
\caption{Pressure dependence of K$\beta$ emission spectra of iron. The spectra have been aligned and normalized to the K$\beta_{1,3}$ mainline. All the spectra measured between 4\,K and 583\,K are shown since there is no temperature dependence (see text and Fig.~\ref{fig:XES_scatter} for details). Inset: zoom of the K$\beta'$ satellite region after subtracting a high-pressure reference and binning the data.}
\label{fig:XES}
\end{figure}
\begin{figure}[!t]
\centering
\includegraphics[width=\linewidth]{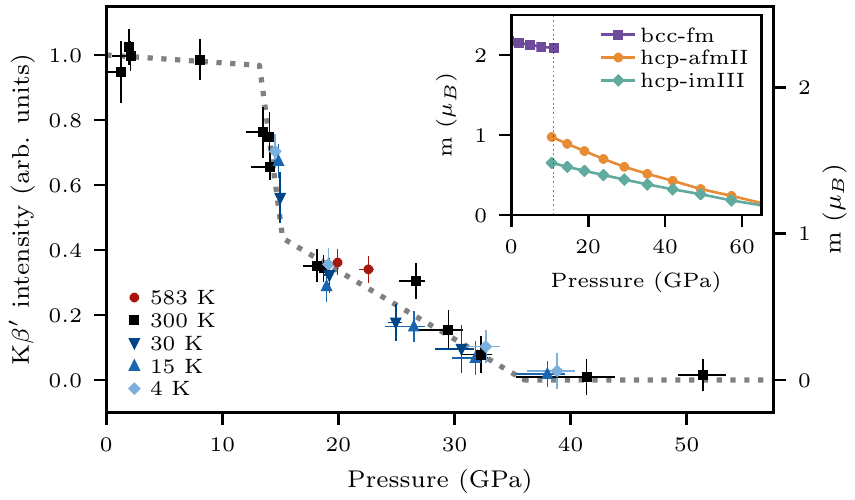}
\caption{Pressure dependence of the K$\beta'$ integrated intensity expressed with respect to a high-pressure reference. The right scale is a linear mapping of K$\beta'$ intensity to magnetic moment using 2.22\,$\mu_B$ at ambient pressure. The gray dashed lines are guides for the eye. Inset: pressure dependence of the average magnetic moment per site found with our DFT calculations. The $\alpha$-$\varepsilon$ transition pressure from a tangent construction is shown as a vertical dashed line.}
\label{fig:XES_scatter}
\end{figure}
%

\section*{Intrinsic magnetic moment}

Hard x-ray photon-in photon-out spectroscopy is well-suited to investigate magnetism in $3d$ compounds under high pressure \cite{Rueff2010, mittelHPmag}. In particular, K$\beta$ x-ray emission spectroscopy (XES) is an established probe of magnetism in iron \cite{Rueff1999a, Rueff2008a, Monza2011} and iron-based compounds \cite{Badro2003, Badro2004, PhysRevLett.98.196404, PhysRevB.97.180503}. K$\beta$ ($3p \rightarrow 1s$) fluorescence has an intense mainline (K$\beta_{1,3}$) and a weaker, low-energy satellite region (K$\beta'$), as shown in our spectra in Fig.~\ref{fig:XES}. This splitting is primarily due to the $3p$-$3d$ exchange interaction between the $3p$ core hole and the majority-spin of the incomplete $3d$ shell in the final state 
\cite{Tsutsumi1959}. 
Therefore, K$\beta$ spectroscopy probes the unpaired $3d$ spin occupation, in other words the $3d$ spin angular momentum. In the case of iron, this corresponds approximately to the local magnetic moment magnitude since the orbital angular momentum is essentially quenched.

We performed K$\beta$ XES on iron over a large range of pressures (0--51.5\,GPa) and temperatures (4--583\,K) using an argon pressure-transmitting medium. Numerous isothermal runs were performed with a monotonically increasing pressure. The spectra are shown in Fig.~\ref{fig:XES} after alignment and normalization to the K$\beta_{1,3}$ mainline ($\approx 7057$\,eV). The relative change in satellite intensity is determined by subtracting a polynomial fit of the highest pressure point (51.5\,GPa, 300\,K) in the K$\beta'$ region and integrating the intensity of the resultant difference spectrum (Fig.~\ref{fig:XES}, inset). 

The pressure dependence of the K$\beta'$ integrated intensity is shown in Fig.~\ref{fig:XES_scatter}. The satellite intensity decline at 15\,GPa corresponds to the $\alpha$-$\varepsilon$ transition, in agreement with x-ray diffraction \cite{doi:10.1029/JB091iB05p04677, Boehler1990} and M\"{o}ssbauer spectroscopy \cite{Taylor1992moss} using an argon pressure-transmitting medium. The use of a more hydrostatic pressure-transmitting medium, coupled with increased statistics and a different analysis technique, shows this transition and the intensity after this transition significantly better than previous XES measurements \cite{Monza2011}. The $\alpha$-$\varepsilon$ transition pressure from previous studies, using an argon pressure medium, support that the signal above 18\,GPa is intrinsic to the $\varepsilon$-iron phase rather than due to a minority $\alpha$-iron phase. The lack of temperature dependence around 20\,GPa supports its intrinsic nature since we would expect a larger K$\beta'$ XES signal from more $\alpha$-iron impurities at lower temperatures due to the increased transition width \cite{Dewaele2017}. Furthermore, a signal due to exclusively $\alpha$-iron impurities implies unphysical values, e.g. 35\% (10\%) $\alpha$-iron at 20\,GPa (30\,GPa). Therefore, we find an intrinsic local magnetic moment in $\varepsilon$-iron that persists until 30--40\,GPa. This is coincidentally the pressure region above which superconductivity disappears in $\varepsilon$-iron. The pressure dependence of the magnetic moment from our first-principles calculations shows a remarkable similarity to our results (Fig.~\ref{fig:XES_scatter}, inset): a linear decrease in $\alpha$-iron, a sharp drop across the transition, and finally a linear decrease in $\varepsilon$-iron with a larger slope. Our calculations predict a null moment above 70\,GPa, which is at a higher pressure than found by XES measurements, nonetheless a naive linear mapping of the K$\beta'$ integrated intensity to magnetic moment (right scale of Fig.~\ref{fig:XES_scatter}) gives $0.74\,\mu_B$ at 20\,GPa, which agrees remarkably well with $0.77\,\mu_B$ calculated for afmII. This magnetic moment also agrees fairly well with our new spin-intensity-modulated phases discussed below, in particular imIII which predicts $0.54\,\mu_B$ at 20\,GPa.

\section*{Spin-smectic state}

%
\begin{figure}[tb]
\centering
\includegraphics[width=\linewidth]{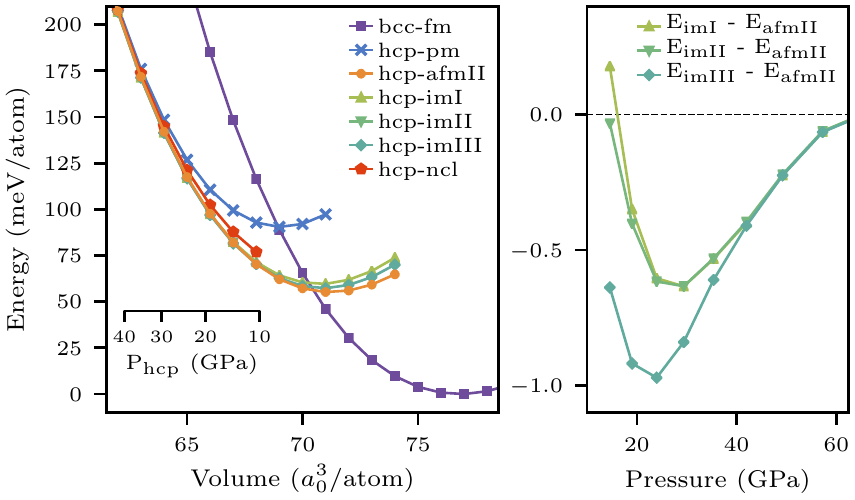}
\caption{Left panel: energy vs. volume per atom for the different phases of iron from DFT calculations at T $=$ 0\,K: ferromagnetic (fm, bcc only), paramagnetic (pm), antiferromagnetic (afm), intensity modulated (im) and non-collinear (ncl), see text for details. The hcp-afmII curve is fit with a Vinet equation of state~\cite{Vinet:1987aa} to determine the pressure used in the right panel and shown inset in the left panel. Right panel: energy difference between the intensity-modulated and afmII phases as a function of pressure at T $=$ 0\,K.}
\label{fig:DFT}
\end{figure}

The ground state determination of $\varepsilon$-iron has proven to be a difficult goal, since the hcp frustration for antiferromagnetism produces a broad range of spin arrangements in a narrow energy window~\cite{cohen2004non,lizarraga2008noncollinear,thakor2003ab}. 
By means of extensive density functional theory (DFT) calculations from first principles at T $=$ 0\,K, we searched for the lowest energy state by perturbing the afmII spin arrangement, the previous best candidate to describe the $\varepsilon$-iron magnetic phase.  Our main finding is that the modulation of the magnetic moment intensity favors the breaking of the afmII symmetry, as reported in Fig.~\ref{fig:DFT}.

\begin{figure}
\centering
\includegraphics[width=\linewidth]{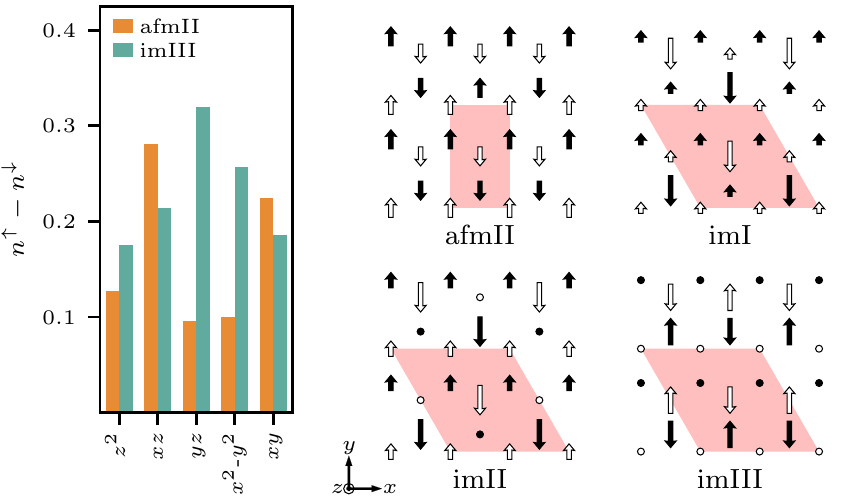}
\caption{Left panel: orbital-resolved local $3d$ polarization of the afmII and imIII phase. The occupation number of the majority ($n^{\uparrow}$) and minority ($n^{\downarrow}$) has been computed by integrating the locally-projected DoS up to the Fermi level. Right panel: magnetic-moment arrangements in the afmII and intensity-modulated phases with the magnetic unit cell shaded in red. White and black arrows are moments belonging respectively to the upper and lower $z$-plane and circles are zero moments. }
\label{fig:DFT_phases}
\end{figure}

We found three different intensity-modulated (im) phases, referred to as imI, imII and imIII (Fig.~\ref{fig:DFT_phases}, right panel). The system is more stable when the magnetic moments' amplitudes are site-dependent ---  all three intensity-modulated phases have a lower energy than afmII from 20~GPa to 60~GPa (Fig.~\ref{fig:DFT}, right panel). The imIII phase is the most extreme case with its alternating magnetic and non-magnetic bilayers and has a lower energy than afmII at all pressures above the $\alpha$-$\varepsilon$ transition. We refer to this distinct arrangement as ``spin-smectic'', in analogy with the smectic phase in liquid crystals, since it also breaks lattice translation invariance. This mechanism is driven by magnetic frustration, yielding the imIII phase as the lowest energy state. Indeed, the intensity-modulated spin patterns can be obtained as local minima or saddle points by considering an isolated tetrahedron, the hcp frustration unit~\cite{auerbach1988,Diep:1992aa}, and minimizing the energy at fixed absolute magnetization. 
Another way of lowering the frustration is to develop noncollinear phases, which are expected in $\varepsilon$-iron~\cite{cohen2004non}. We searched for noncollinear phases in our \emph{ab initio} investigation. The most stable one is reported in Fig.~\ref{fig:DFT} as hcp-ncl and in the SI Appendix (Fig. S4). It is still higher in energy than afmII, suggesting that longitudinal smectic fluctuations are the optimal spin arrangement in the frustrated hcp-iron lattice.  

\begin{figure*}
\centering
\includegraphics[width=\textwidth]{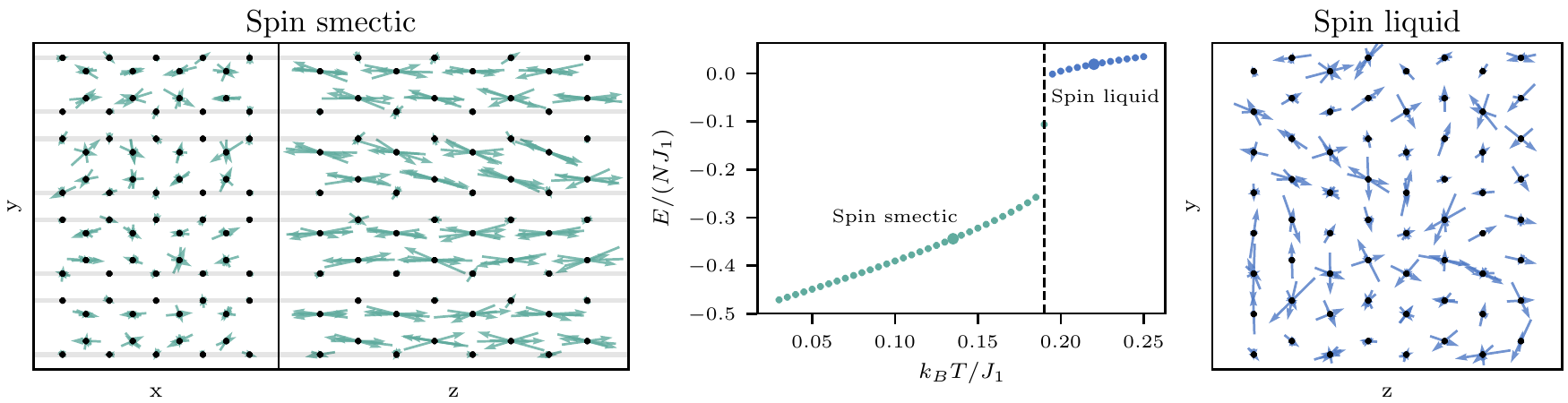}
\caption{Monte Carlo energy per spin (middle panel) as a function of the temperature at $p\approx 24$\,GPa. A representative snapshot of the smectic phase taken at $k_{B}T/J_1=0.135$ is reported in the left panel, the ordered bilayers lying on the $xz$ plane. On the right panel a snapshot of the paramagnetic phase at $k_{B}T/J_1 =0.22$ is shown. The temperature ramp has been performed by cooling the system from a disordered state in thermal equilibrium. For each temperature the canonical distribution has been sampled  by proposing $10^5$ lattice updates for each of the 128 random walkers.}
\label{fig:MC}
\end{figure*}

In the left panel of Fig.~\ref{fig:DFT_phases}, we compare the local spin polarization of the afmII and imIII phases at 19\,GPa (their average local magnetic moment $m$ is instead reported in the inset of Fig.~\ref{fig:XES_scatter}) The homogeneous distribution of spin polarization among the $3d$ orbitals in imIII signals a very weak crystal-field splitting and a large on-site spin degeneracy. This leads to enhanced spin-intensity modulations dictated by the intersite exchange interaction. Thus, we can map the DFT energies onto a generalized Heisenberg model, where the classical spins are allowed to change in both amplitude and direction~\cite{rosengaard1997,ruban2007,shallcross2005}, to account for both transverse and longitudinal spin fluctuations. In this downfolding procedure, we assumed a perfect hcp lattice for the model, since the DFT spin arrangement does not show any anisotropy due to a nearly ideal hcp $c/a$ ratio.

We find that the antiferromagnetic nearest-neighbor Heisenberg model is able to capture the key DFT features, i.e.\ the instability towards spin-intensity modulated phases, and their \emph{ab initio} energy ordering ($E_{\rm imIII}<E_{\rm imII}<E_{\rm imI}<E_{\rm afmII}$) in the $20$--$30$\,GPa region. 
However, in order to properly take into account the magnetic itinerancy of the system, we extended this 
simple
model including local, nearest-neighbour and next-nearest-neighbour interactions up to the $4^{\rm th}$ order in the magnetic moment. Our 
extended
model is able to capture not only the energy hierarchy of the DFT collinear and noncollinear phases, but also to reproduce their average magnetic moment.
The fitting procedure and the evolution of the coefficients as a function of pressure are reported in the SI Appendix(Sec.~3, Tab.~S1, and Fig.~S5).

We investigated the finite-temperature behavior of magnetism in $\varepsilon$-iron by performing classical Monte Carlo (MC) simulations of our generalized Heisenberg model. We observe a first-order phase transition with the critical temperature $T_m$ dropping from $55\,K$ at 20\,GPa to $15\,K$ at 35\,GPa. 
We warn however that the computed critical temperature is not to be considered in a stringent quantitative way, due to the open issue of phase-space sampling in the case of longitudinal fluctuations~\cite{khmelevskyi2018,wysocki2008}; however, its pressure evolution yields the predicted scaling behavior (Fig.~\ref{fig:phase}).

For $T<T_{\rm m}$ the system acquires a spin arrangement of imIII-type, which consists of antiferromagnetically ordered bilayers, interleaved with null magnetic-moment bilayers (see Fig.~\ref{fig:DFT_phases}; Fig.~\ref{fig:MC}, left panel). This particular smectic pattern completely removes the antiferromagnetic frustration generated by the nearest-neighbors $J_1$ interaction in the hcp lattice, since each spin is left with 4 non-zero neighbors of opposite orientation. We note that the DFT energy is minimized by such an arrangement, supporting the validity of our spin model.
Once the smectic order has set in after the transition, the interlayer interaction in the extended Heisenberg model is mediated by the next-nearest-neighbors $J_2$ coupling, whose intensity 
decreases with pressure faster than $J_1$. We remark that in case of a vanishing $J_2$ coupling, the resulting smectic arrangement cannot be detected by neutron scattering because of its lack of interlayer ordering. 

\begin{figure}[t]
\centering
\includegraphics[width=\linewidth]{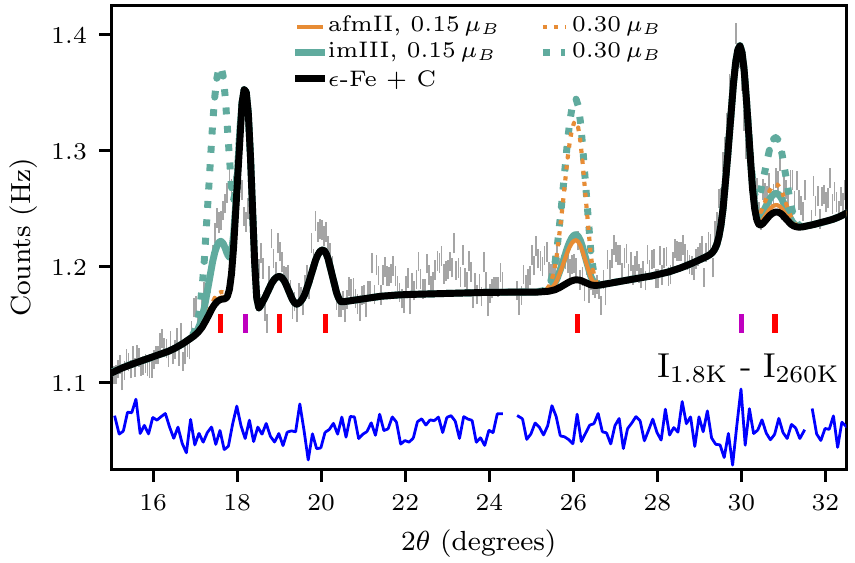}
\caption{Neutron powder diffraction pattern of $\varepsilon$-iron at 20.2\,GPa/1.8\,K performed with $\lambda$ = 1.3\,\AA{} neutrons. The diffraction patterns (gray errors bars) are shown with their Rietveld refinement (black line), where all visible peaks are due to secondary reflections from $\lambda/2$ contamination. The magenta and red ticks indicate reflections from diamond and $\varepsilon$-iron respectively. Simulated diffraction patterns are shown for afmII (orange) and imIII (green-blue) using an average magnetic moment per site of 0.15\,$\mu_B$ (solid line) and 0.3\,$\mu_B$ (dashed line). The binned difference pattern with an 18.5\,GPa/260\,K reference is shown below using the same scale.}
\label{fig:NPD_zoom}
\end{figure}
%

\section*{Unobserved static magnetic order}
We investigated possible magnetic ordering in $\varepsilon$-iron using neutron powder diffraction. This technique is particularly suited to measuring antiferromagnetic structures since their supercells imply certain magnetic reflections away from the nuclear reflections and towards lower scattering angles $2\theta$ where the magnetic form factor is greatest. Using the techniques described in Ref.~\cite{Klotz-HPR-NPD-Fe}, we measured $\varepsilon$-iron above 20\,GPa and down to 1.8\,K. We see a complete $\alpha$-$\varepsilon$ transition during a quasi-isothermal pressure ramp to 18.5\,GPa and a slight pressure increase to 20.2\,GPa upon cooling to 1.8\,K.

The low-$2\theta$ range of the diffraction pattern at 20.2\,GPa/1.8\,K is shown in Fig.~\ref{fig:NPD_zoom} as gray error bars. The whole-pattern Rietveld refinement with $\varepsilon$-iron and diamond is shown as a black line. All the nuclear reflections in this region are actually weak secondary reflections of the sample and sintered diamond anvils due to $\lambda/2$ contamination (0.2\%). The two large peaks around $18.2^{\circ}$ and $30.0^{\circ}$ are diamond (111) and (220) secondary reflections. Secondary reflections from $\varepsilon$-iron are found at $17.6^{\circ}$ (100), $19.0^{\circ}$ (002), $20.1^{\circ}$ (101), $26.1^{\circ}$ (102), and $30.8^{\circ}$ (110). No magnetic peaks are clearly visible in $\varepsilon$-iron. The difference pattern with a high-temperature reference (18.5\,GPa/260\,K) is shown as the blue line. There are no hints of magnetic order appearing from 260\,K to 1.8\,K and the background features are temperature independent.

We simulated magnetic diffraction patterns for afmII and imIII with the goal to establish the upper limits of their magnetic moments consistent with the diffraction pattern. The upper limits are expressed as the average ordered magnetic moment per site. For afmII this corresponds to the same spin intensity at each site; however, for imIII this corresponds to half the spin intensity of the non-zero sites since the other sites have zero spin. The upper limit at 20.2\,GPa/1.8\,K is estimated to be 0.15\,$\mu_B$ for both afmII and imIII, and their simulations are shown as solid colored lines. Simulations with twice the upper limit (0.3\,$\mu_B$) are shown as dashed colored lines to give a sense of scale. 

The upper limit on the magnetic moment of the afmII phase is five times less than 0.77\,$\mu_B$ predicted by DFT. The measurement was performed at 1.8\,K, well below the predicted ordering temperature of 75\,K at 21\,GPa for afmII \cite{Papandrew2006} and 69\,K at 16\,GPa for an incommensurate antiferromagnetic structure \cite{Thakor2003}. Furthermore, there is no noticeable change with temperature (1.8--260\,K) in the diffraction pattern. Therefore, if $\varepsilon$-iron hosts afmII long-range magnetic order then its moment is in very strong disagreement with our estimate from both XES and DFT.

The imIII phase is also undetected by NPD and has an upper limit more than three times less than 0.54\,$\mu_B$ predicted by DFT. We computed an ordering temperature of 55\,K at 20\,GPa for the spin-smectic, imIII-like state (Fig.~\ref{fig:phase} and SI Appendix). Like afmII, our NPD results are also incompatible with a fully ordered imIII state (albeit less so), however they are compatible with an imIII-like state without interlayer ordering.

\section*{Discussion}
We have found evidence of local magnetic moments without any long-range magnetic order in $\varepsilon$-iron. Previous DFT calculations predicted that the afmII configuration is the ground state of $\varepsilon$-iron 
\cite{Steinle-Neumann2004}. Using neutron powder diffraction we give an upper bound on the afmII moment more than five times less than computed with DFT for afmII or estimated from our experimental XES results. These results are consistent with other reports against long-range afmII order \cite{Papandrew2006, Goncharov2003}.

Our DFT calculations found spin-intensity-modulated phases which are lower in energy than the afmII state. Among these, the imIII state achieves the lowest energy by coping with the antiferromagnetic frustration of the hcp lattice with its smectic bilayer structure, where each atom in the afm bilayers is antiferromagnetically coupled to four nearest neighbors. 
DFT ---~and consequently derived spin models~--- tends to overestimate spin order in iron-based materials with significant itinerancy~\cite{mazin2009key}, therefore a similar effect is also expected in $\varepsilon$-iron. Nevertheless, we found that the ferromagnetic interlayer coupling between the afm bilayers weakens rapidly with pressure, leading to stronger spin fluctuations around an imIII-like smectic pattern.

The smectic spin fluctuations which prevent long-range order can also be triggered by the im phases near-degeneracy, in a scenario similar to the one proposed to explain nematicity in FeSe~\cite{glasbrenner2015}.
Both scenarios are compatible with our null neutron powder diffraction results. As well, an imIII-like state could be undetectable by M\"{o}ssbauer spectroscopy if the dynamics of the bilayers is faster than the $\approx$100\,ns timescale of M\"{o}ssbauer spectroscopy. Ferromagnetic spin fluctuations are favored by transport measurements which report non-Fermi liquid behavior with $T^{5/3}$ dependence in this
pressure region 
\cite{Holmes2004, Yadav2013}. 
Spin fluctuations are also inferred from LDA+DMFT calculations due to an underestimation of resistivity, since spatial spin correlations cannot be captured within this framework. Moreover, the LDA+DMFT paramagnetic equation of state remarkably matches the experimental one above 40\,GPa, but shows an appreciable divergence precisely in the 15--40\,GPa region ~\cite{Pourovskii2014}.

We have shown with XES that the local magnetic moment in $\varepsilon$-iron disappears at 30--40\,GPa. The temperature below which resistivity measurements find a $T^{5/3}$ dependence is known as $T^{\rm *}$ and it also approaches zero in this pressure range \cite{Yadav2013}. This is also the pressure where LDA+DMFT infers that spin fluctuations no longer play an important role \cite{Pourovskii2014}. An anomalous Debye sound velocity, $c/a$ ratio \cite{Ono2010}, and M\"{o}ssbauer center shift around 40\,GPa have also been reported and were attributed to an electronic topological transition \cite{Glazyrin2013}. However, x-ray diffraction results recently found no evidence of this electronic topological transition \cite{Dewaele2017} and the most likely origin is the loss of magnetism we find in this study. The disappearance of magnetism around 30--40\,GPa in $\varepsilon$-iron means it does not play an important geophysical role in the Earth \cite{Saxena2001}, however it could still play a role in smaller rocky planets such as Mercury and exoplanets. 
Furthermore, measurements of $\varepsilon$-iron below 40\,GPa should not be extrapolated to higher pressures due to the effects of magnetism in this low pressure region. The calculated magnetic moments in imIII only disappear above 70\,GPa (Fig.~\ref{fig:XES_scatter}, inset), which is higher than the experimental results discussed above. This discrepancy between experiment and calculations is likely due to the well-known overestimation of magnetic moments when using mean-field calculations such as DFT.

The disappearance of superconductivity \cite{Shimizu2001, Yadav2013} coincides with the 30--40\,GPa region discussed above and warrants speculation about its connection with the spin-smectic state (Fig.~\ref{fig:phase}). The role of spin fluctuations was discussed shortly after the discovery of superconductivity in $\varepsilon$-iron because of the failure of conventional phonon-mediated Bardeen–Cooper–Schrieffer theory \cite{Bose2003, Mazin2002, Jarlborg2002}. These attempts were quickly abandoned since they required complicated competition between many interactions and were unable to replicate the relatively small pressure range. We believe that this line of inquiry should be revisited in the context of a spin-smectic state. Furthermore, previous studies used DFT calculations as input which predict the disappearance of magnetism at higher pressures than we experimentally report here. 
\section*{Conclusions}

We have used K$\beta$ x-ray emission spectroscopy to reveal the existence of an intrinsic local magnetic moment in $\varepsilon$-iron. Our neutron powder diffraction results found no magnetic order and gave upper limits which suggest that the previously proposed ground state, afmII, does not order. Our DFT calculations found spin-intensity-modulated (im) phases lower in energy than the afmII phase. This spin-intensity modulation reduces the effective frustration with a perfect cancellation found in the imIII arrangement, which also is the lowest energy state in the 15--35\,GPa pressure range. Based on these results, we derived an extended antiferromagnetic Heisenberg Hamiltonian which correctly reproduces the DFT energy and magnetization hierarchy. MC simulations showed that the im-type arrangements survive at low temperature, suggesting that the long-range magnetic order is hampered by spin-smectic fluctuations in the low-temperature range. The spin-smectic state is compatible with our experimental findings but is also particularly elusive to detection. Muon spectroscopy would be very informative and our results are motivation to push the current pressure limitations of this technique which are currently an order of magnitude too small \cite{muon}.
\section*{Materials and Methods}

Experimental (blair.lebert@gmail.com) and theoretical (gornitom@gmail.com) data is available upon request. XES performed on GALAXIES at SOLEIL \cite{GALAXIES, GALAXIES2} using 1-m Rowland circle in transmission geometry with Si(531) analyzer with a 30 $\times$ 80\,$\mu$m$^2$ beam of 9\,keV. 5\,$\mu$m-thick Fe foils loaded in Rh gasket of DAC with argon and ruby or SrB$_4$O$_7$:Sm$^{2+}$. NPD performed \cite{D20_2015} on D20 \cite{Hansen} at ILL using technique reported in Ref.~\cite{Klotz-HPR-NPD-Fe}. Rietveld refinement and simulations performed with \textsc{fullprof} \cite{FULLPROF} using Fe$^{3+}$ magnetic form factor. DFT used pw.x code of \textsc{Quantum ESPRESSO} \cite{Giannozzi:2017} in the PAW scheme~\cite{Blochl:1994aa} with $3s$ and $3p$ electrons in valence and using the GGA approximation with the PBE functional~\cite{Perdew:1996ab}. Plane-wave (density) cutoff set to 100\,Ry (400\,Ry). BZ integrated on $24 \times 24 \times 24$ $\mathbf{k}$-mesh (8-atom unit cell) with $0.25$\,mRy Gaussian broadening. The Vinet equation of state~\cite{Vinet:1987aa} has been used to determine the pressure in the $\varepsilon$-phase, with the best-fit parameters $V_0 = 71.183\,a_0^3$, $K_0 = 199.76$\,GPa and $K_0' = 6.2$. MC was used to study our generalized Heisenberg model (SI Appendix). We adopted a multi-walker approach, running 128 independent MC samplings per temperature in parallel, each of them performing $10^5$ lattice sweeps.To assess the convergence of thermodynamical averages, we performed finite-size scaling up to $16\times 16 \times 10$ hcp unit cells.

\section*{Acknowledgements}

Beamtime from SOLEIL (20130318, 20120694) and ILL \cite{D20_2015}. B.W.L \& T.G. supported by the French state funds managed by the ANR within the ``Investissements d’Avenir'' programme under reference ANR-11-IDEX-0004-02, and within the framework of the Cluster of Excellence MATISSE led by Sorbonne Universit\'{e}.

\bibliographystyle{unsrt}
\bibliography{Fe-bib}

\end{document}